\documentclass[a4paper,11pt]{article}
\usepackage{pos}

\title{Top quark mass and properties in ATLAS and CMS}

\author*[1]{Sebastian Wuchterl}

\affiliation{Deutsches Elektronen-Synchrotron (DESY),\\
  Notkestrasse 85, Hamburg, Germany}
\note{For the ATLAS and CMS Collaborations.}
\emailAdd{sebastian.wuchterl@cern.ch}

\abstract{
For standard model (SM) processes involving top quarks, such as top quark-antiquark pair production or single top quark production, the theoretical predictions depend on fundamental parameters of the SM like the top quark mass.
Using the large data sample collected at the CERN LHC in the second data-taking period by the ATLAS and CMS experiments, these parameters can be extracted in experimental measurements with high precision.
Furthermore, properties of the production processes such as quark polarization or asymmetries can be measured.
Exploiting the large luminosity of the data set, new kinematic regimes and methods are explored, as for events in which the top quarks have a very large transverse momentum.
In these proceedings, recent measurements of SM parameters and top quark properties are presented.
Individual results are also interpreted in terms of effective field theory extensions of the SM.
}

\FullConference{%
  The Tenth Annual Conference on Large Hadron Collider Physics - LHCP2022\\
  16-20 May 2022\\
  online
}

\newcommand{\ttbar}{\ensuremath{\mathrm{t\bar{t}}}}
\newcommand{\ttbarjet}{\ensuremath{\mathrm{t\bar{t}+jet}}}
\newcommand{\fbinv}{\ensuremath{\mathrm{fb^{-1}}}}
\newcommand{\mttbar}{\ensuremath{m_\mathrm{{t\bar{t}}}}}
\newcommand{\mt}{\ensuremath{\mathrm{m_{t}}}}
\newcommand{\mtpole}{\ensuremath{\mathrm{m_{t}^{pole}}}}
\newcommand{\mtmsr}{\ensuremath{\mathrm{m_{t}^{MSR}}}}
\newcommand{\mtmc}{\ensuremath{m_\mathrm{{t}}^\mathrm{{MC}}}}

\begin{document}
\renewcommand{\logo}{\relax}
\maketitle

\section{Introduction}
At the CERN LHC, top quarks are produced with a high production rate, predominantly in top quark-antiquark pair (\ttbar) production via the gluon fusion mechanism.
However, given the data set available from the second data taking period (LHC Run II) (2015-2018) at a center-of-mass-energy of 13\,TeV, also processes like single top \textit{t}-channel production can be studied with high precision.
The large mass of the top quark, \mt, is a free parameter of the SM and needs to be determined experimentally. Its large value hints that the top quark plays a special role within the standard model (SM), especially for the electroweak symmetry breaking.
Therefore, the precise study of processes involving top quarks can shed a light also on beyond the SM physics.
Further, theoretical predictions for top quark processes depend on \mt\ or other fundamental SM parameters, allowing for their extraction from measurements of production cross sections or kinematic observables.
Recent measurements performed by the ATLAS~\cite{bib:ATLAS} and CMS~\cite{bib:CMS} Collaborations are presented here in these proceedings.

\section{Top quark mass measurements}
Measurements of \mt\ can be distinguished depending on the experimental procedures employed.
On one hand, \mt\ can be extracted in a well-defined theoretical renormalization scheme, e.g., pole or MSR~\cite{bib:MSR}, by comparing absolute or differential cross section measurements to theoretical predictions at fixed-order. This approach is usually referred to as indirect measurement.
On the other hand, \mt\ can be measured by comparing multi-purpose Monte Carlo (MC) predictions to variables sensitive to the reconstructed energy of the top quarks. Measurements of this type are commonly denoted as direct measurements, and the value measured can be referred to as the top quark MC mass \mtmc.
They lack a clear theoretical interpretation compared to indirect measurements because of the modeling of non-perturbative effects in the MC. Hence, an interpretation uncertainty on the order of 0.5--1 GeV reflecting the usage of probabilistic MC generators~\cite{bib:hoangmass,bib:massTheoNason} is added.

The most precise direct measurement for \mt\ to date was performed recently by the CMS Collaboration using $35.9$\,\fbinv\ of pp collision data~\cite{CMS-PAS-TOP-20-008}. Events in the \ttbar\ decay channel with one lepton are analyzed, and a kinematic fit is performed to reconstruct the top quarks. Using a profiled likelihood fit in 5 dimensions, the top quark mass is measured. The resulting value is $\mt=171.77\pm0.38\,\mathrm{GeV}$, which is in good agreement with previous measurements and improves the precision by $0.12\,\mathrm{GeV}$.

\begin{figure}[htbp]
\centering
\includegraphics[width=0.49\textwidth]{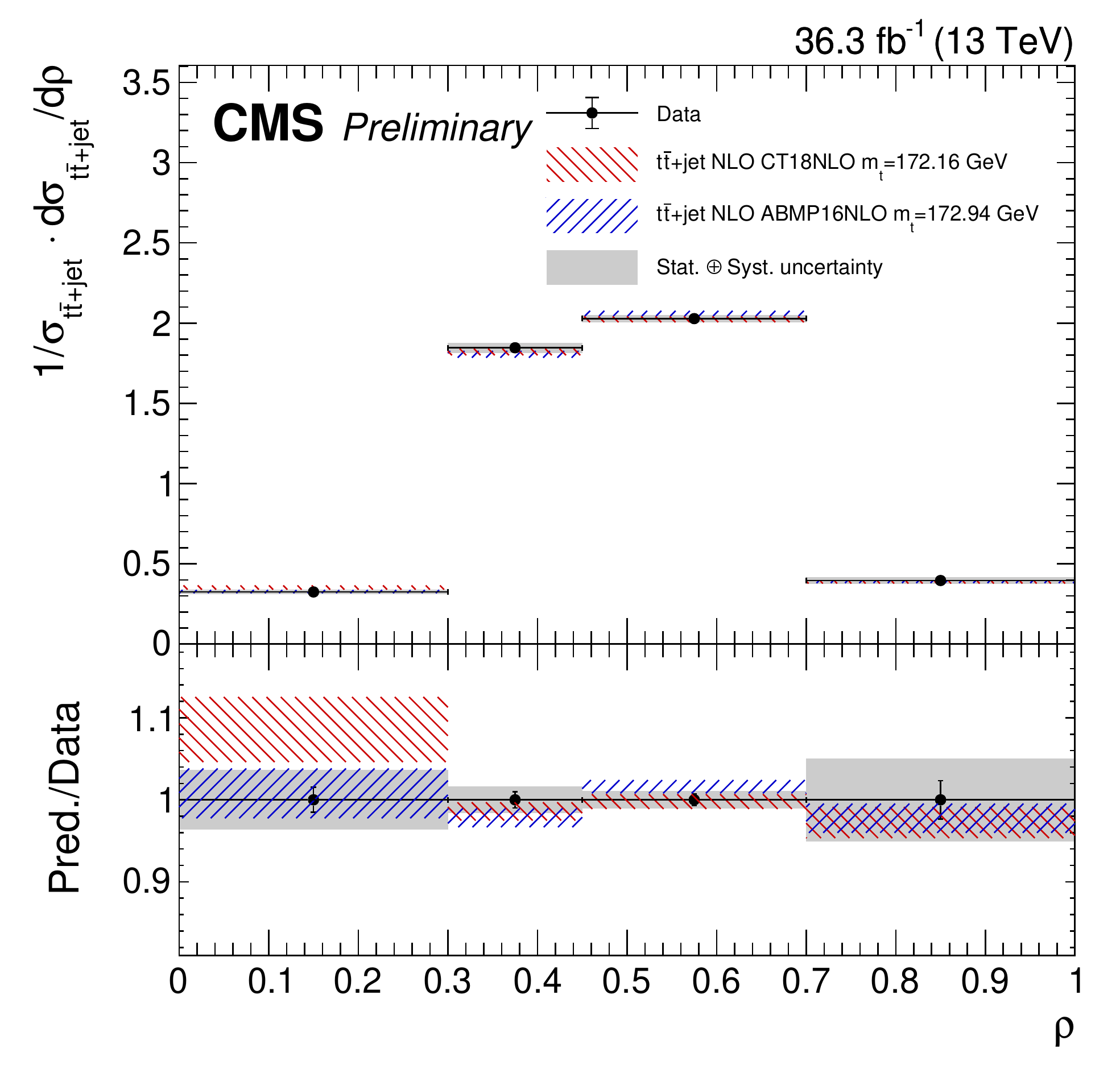}
\includegraphics[width=0.503\textwidth]{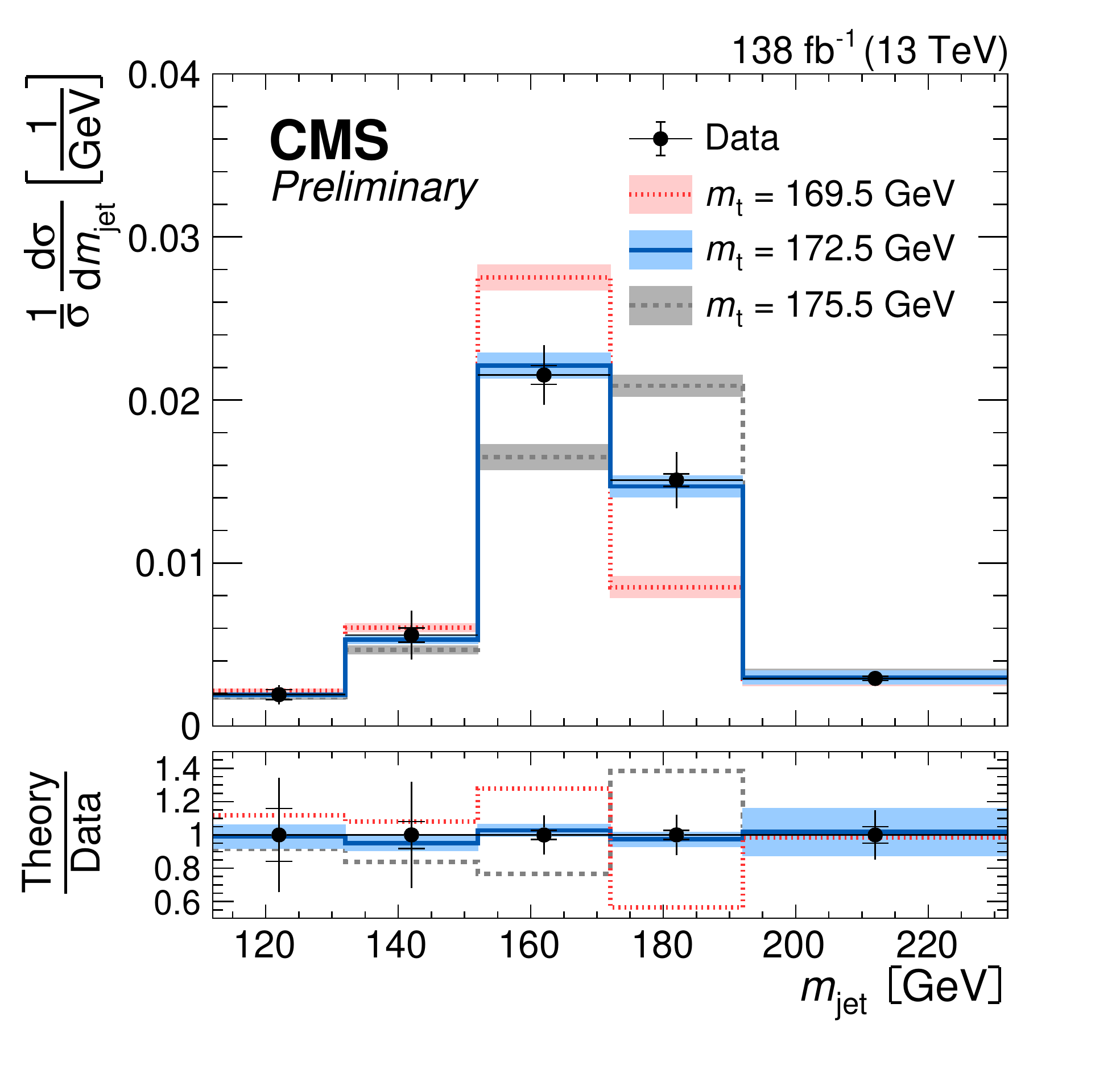}
\caption{Measured cross section as a function of the $\rho$ observable compared to theoretical predictions for best-fit mass values (left)~\cite{CMS-PAS-TOP-21-008}. Unfolded cross section for \ttbar\ production as a function of the boosted jet mass compared to MC predictions for varying top quark masses (right)~\cite{CMS-PAS-TOP-21-012}.}
\label{fig:mass1}
\end{figure}

The pole mass \mtpole\ was also measured by CMS, exploiting the mass sensitivity of \ttbar\ production with at least one additional jet (\ttbarjet)~\cite{CMS-PAS-TOP-21-008}. The normalized differential cross section as a function of the $\rho$ observable defined as $340\,\mathrm{GeV}/m_{\ttbarjet}$ is measured at the parton level using a profiled likelihood unfolding approach. Machine learning methods are used for the reconstruction of $\rho$ and the event classification. From a comparison to next-to-leading-order (NLO) predictions~\cite{bib:ttjPheno}, \mtpole is extracted using different parton distribution functions. Using ABMP16NLO~\cite{bib:ABMP16}, \mtpole\ is measured to be $172.94\pm1.37\,\mathrm{GeV}$.
The best-fit predictions and the measurement are shown in Fig.~\ref{fig:mass1} (left).

The connection between \mtpole\ and \mtmc\ can be probed in the boosted regime by measuring the invariant mass of the boosted jets. Here, a measurement by CMS using a data set corresponding to 138\,\fbinv~\cite{CMS-PAS-TOP-21-012} is presented for the first time. The differential cross section is measured as a function of the invariant jet mass and is unfolded to particle level. Comparing it to MC predictions, \mtmc\ is determined to be $172.76\pm0.81\,\mathrm{GeV}$. The unfolded distribution is shown in Fig.~\ref{fig:mass1} (right) compared to predictions for different values of \mt. With respect to previous measurements, the result improves by a factor of two. This is mainly due to dedicated calibrations of the jet mass scale and tuning of the final state radiation scale in the MC.

Using the same observable and theoretical predictions at next-to-leading-log precision, the relation between \mtmc\ and \mtpole and \mtmsr is studied by the ATLAS Collaboration~\cite{ATL-PHYS-PUB-2021-034}. Template fits for the theoretical prediction to the simulation at particle level are performed while the numerical difference between \mtmc\ and \mtpole\ or \mtmsr is determined.
The difference between \mtmc\ and $\mtmsr(R=1\,\mathrm{GeV})$ is measured to be $80^{+350}_{-410}\,\mathrm{MeV}$, and $350^{+300}_{-360}\,\mathrm{MeV}$ for \mtmc\ and \mtpole. A scale of $R=1\,\mathrm{GeV}$ is used due to the numerical similarity to the pole mass ($\mtmsr(R=1\,\mathrm{GeV}) \approx \mtpole$).

\section{Measurements of top quark properties}
Given the special features of top quarks, the polarization of top quarks and antiquarks in \textit{t}-channel single top production can be investigated, as done by ATLAS using 139\,\fbinv\ of pp collision data~\cite{ATLAS:2022vym}. The polarization vectors for t and $\mathrm{\bar{t}}$ are determined from a likelihood fit to variables that are constructed to yield maximum sensitivity. All six components of the polarization vectors are in good agreement with the SM expectation at next-to-NLO accuracy.
Additionally, distributions of the angles of the lepton in the top quark rest frame are unfolded to particle level and effective-field-theory (EFT) couplings affecting the tWb vertex are probed.

\begin{figure}[htbp]
\centering
\includegraphics[width=0.49\textwidth]{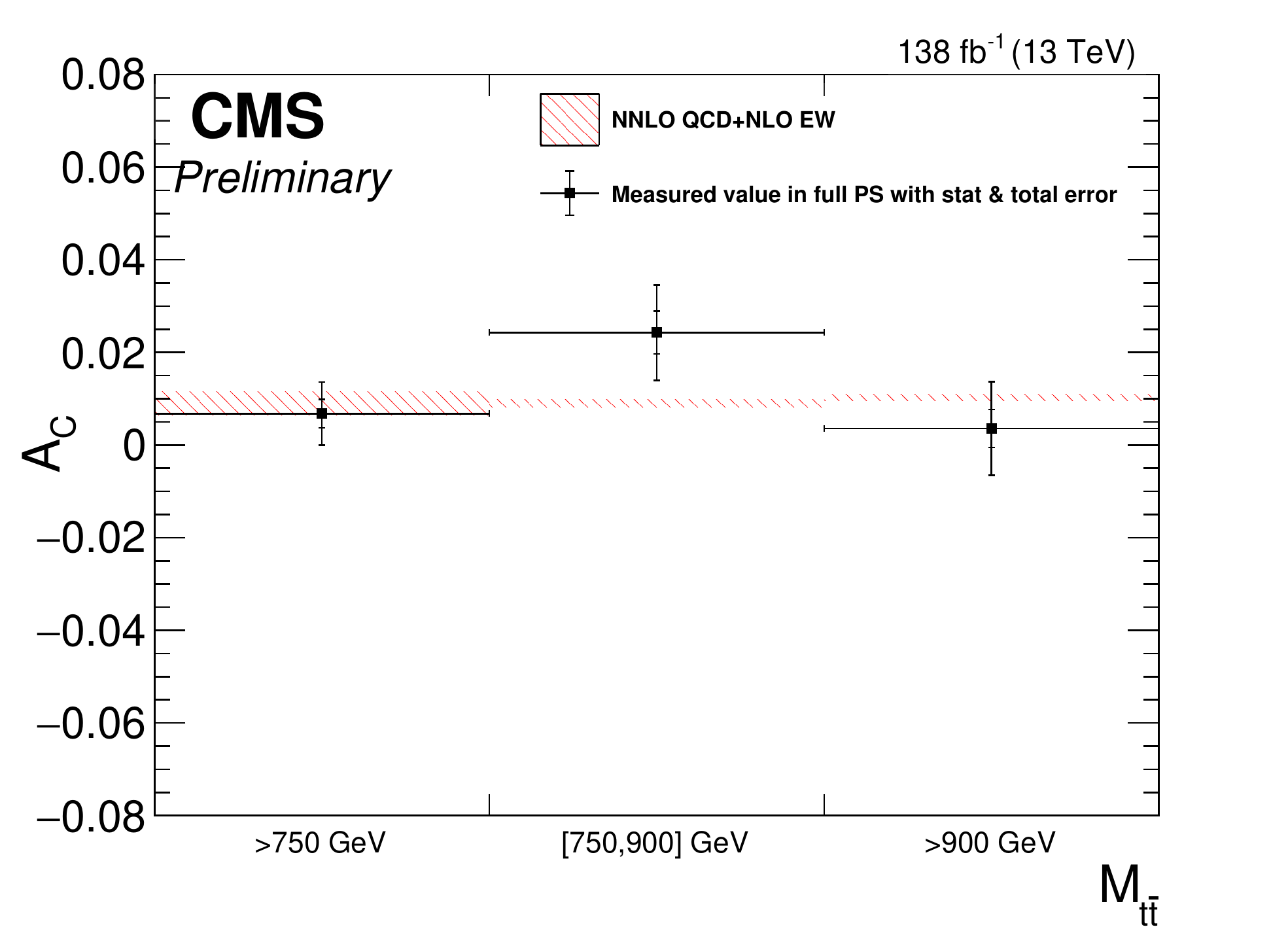}
\includegraphics[width=0.49\textwidth]{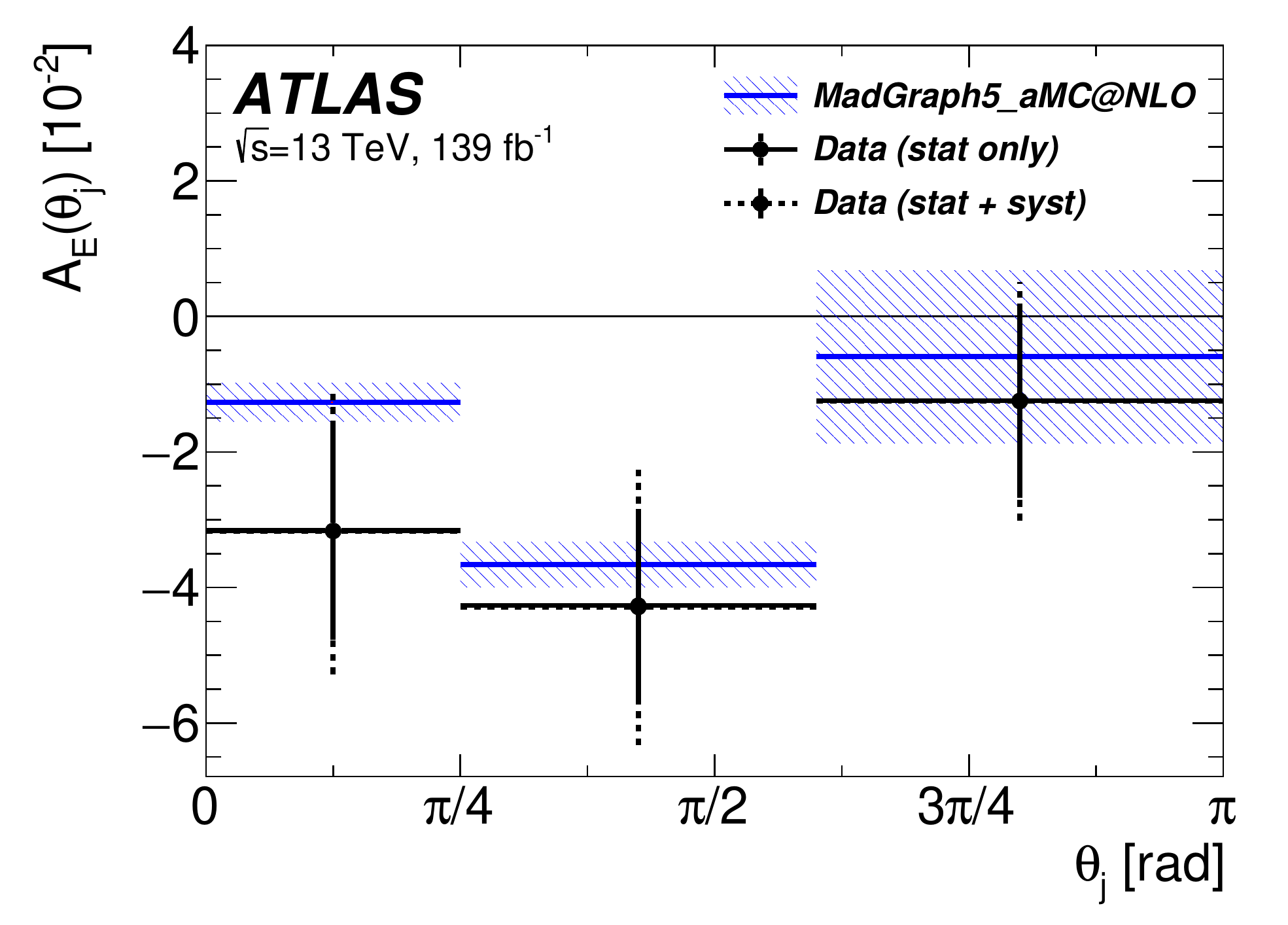}
\caption{Charge asymmetry in the full phase space compared to theoretical predictions(left)~\cite{CMS-PAS-TOP-21-014}. Energy asymmetry for \ttbarjet\ production in bins of the scattering angle of the additional jet (right)~\cite{ATLAS:2021dqb}.}
\label{fig:asymm}
\end{figure}

Using 138\,\fbinv\ of pp collision data, CMS measured the \ttbar\ charge asymmetry $A_C$ in boosted top quark pair events with one lepton in the final state~\cite{CMS-PAS-TOP-21-014}. The measurement is presented here for the first time. The charge asymmetry is predicted to be zero ($\approx6.6\%$) in the SM for the gluon fusion (quark-antiquark) production mechanism. Beyond-the-SM effects could lead to a non-negligible change of the value of $A_C$. It is the first measurement of $A_C$ analyzing pp collision data at $\sqrt{s}=13\,\mathrm{TeV}$ with highly-boosted \ttbar\ events for invariant masses of $\mttbar>750\,\mathrm{GeV}$.
The value for $A_C$ is measured in three bins of \mttbar\ in the fiducial and full phase space using a likelihood unfolding method.
The measured data in the full phase space compared to theoretical predictions is shown in Fig.~\ref{fig:asymm} (left).
A good agreement between measurement and SM expectation is observed.

Similarly, the ATLAS Collaboration measured the energy asymmetry in \ttbarjet\ production~\cite{ATLAS:2021dqb}, which yields an alternative method to determine the charge asymmetry. The measurement is performed at particle level as a function of the scattering angle of the additional jet using a likelihood unfolding method. Also, here, events with boosted top quark jets are analyzed. The result is shown in Fig.~\ref{fig:asymm} (right). The energy asymmetry is measured to be $0.043\pm0.020$, which is in good agreement with the SM prediction. Additionally, the result is interpreted in the context of EFT, setting limits on four-quark operators.

\section{Summary}
In this contribution, recent measurements of the top quark mass and standard model (SM) properties are presented.
Significant progress was made by the ATLAS and CMS Collaborations using the data collected during LHC Run II at a center of mass energy of $13\,\mathrm{TeV}$, allowing to analyze phase space regions with unprecedented precision for example in the boosted regime of top quark-antiquark pair production (\ttbar).
Multiple top quark mass measurements by CMS and an interpretation study by ATLAS are presented, improving the understanding of the relation between well-defined mass definitions and direct measurements.
Both collaborations also measured SM properties for top quark production processes, such as the charge and energy asymmetry in \ttbar\ or \ttbar\ production with one additional jet, respectively.
Unfolded results are in good agreement with SM predictions, and interpretations are performed in terms of effective-field-theory (EFT) extensions of the SM.
Finally, the top quark and antiquark polarizations are measured by ATLAS in \textit{t}-channel single top production events, yielding results that confirm the SM expectation. Also here EFT coupling limits are set, improving with respect to previous measurements.

\bibliographystyle{JHEP}
\bibliography{main}

\providecommand{\href}[2]{#2}\begingroup\raggedright\begin{thebibliography}{10}

\bibitem{bib:ATLAS}
{ATLAS Collaboration}, \emph{The {ATLAS} experiment at the {CERN} {L}arge
  {H}adron {C}ollider},
  \href{https://doi.org/10.1088/1748-0221/3/08/s08003}{\emph{JINST} {\bfseries
  3} (2008) S08003}.

\bibitem{bib:CMS}
{CMS Collaboration}, \emph{{The CMS Experiment at the CERN LHC}},
  \href{https://doi.org/10.1088/1748-0221/3/08/S08004}{\emph{JINST} {\bfseries
  3} (2008) S08004}.

\bibitem{bib:MSR}
A.H.~Hoang, A.~Jain, I.~Scimemi and I.W.~Stewart, \emph{Infrared
  renormalization-group flow for heavy-quark masses},
  \href{https://doi.org/10.1103/PhysRevLett.101.151602}{\emph{Phys. Rev. Lett.}
  {\bfseries 101} (2008) 151602}.

\bibitem{bib:hoangmass}
A.H.~Hoang, \emph{What is the top quark mass?},
  \href{https://doi.org/10.1146/annurev-nucl-101918-023530}{\emph{Ann. Rev.
  Nucl. Part. Sci.} {\bfseries 70} (2020) 225}
  [\href{https://arxiv.org/abs/2004.12915}{{\ttfamily 2004.12915}}].

\bibitem{bib:massTheoNason}
S.~Ferrario~Ravasio, T.~Je\v{z}o, P.~Nason and C.~Oleari, \emph{{A theoretical
  study of top-mass measurements at the LHC using NLO+PS generators of
  increasing accuracy}},
  \href{https://doi.org/10.1140/epjc/s10052-019-7336-9}{\emph{Eur. Phys. J. C}
  {\bfseries 78} (2018) 458}
  [\href{https://arxiv.org/abs/1906.09166}{{\ttfamily 1906.09166}}].

\bibitem{CMS-PAS-TOP-20-008}
{CMS Collaboration}, \emph{{A profile likelihood approach to measure the top
  quark mass in the lepton+jets channel at $\sqrt{s}=13~\mathrm{TeV}$}},  CMS
  Physics Analysis Summary
  \href{https://cds.cern.ch/record/2806509}{CMS-PAS-TOP-20-008}, CERN, Geneva
  (2022).

\bibitem{CMS-PAS-TOP-21-008}
{CMS Collaboration}, \emph{{Measurement of the top quark pole mass using
  $\text{t}\overline{\text{t}}\text{+jet}$ events in the dilepton final state
  at $\sqrt{s}=$ 13 TeV}},  CMS Physics Analysis Summary
  \href{https://cds.cern.ch/record/2804936}{CMS-PAS-TOP-21-008}, CERN, Geneva
  (2022).

\bibitem{bib:ttjPheno}
S.~Alioli, J.~Fuster, M.V.~Garzelli, A.~Gavardi, A.~Irles, D.~Melini et~al.,
  \emph{Phenomenology of $\ttbar\mathrm{j}+\mathrm{X}$ production at the
  {LHC}}, \href{https://doi.org/10.1007/JHEP05(2022)146}{\emph{JHEP} {\bfseries
  05} (2022) 146} [\href{https://arxiv.org/abs/2202.07975}{{\ttfamily
  2202.07975}}].

\bibitem{bib:ABMP16}
S.~Alekhin, J.~Bl{\"u}mlein and S.~Moch, \emph{{NLO PDFs} from the {ABMP16
  fit}}, \href{https://doi.org/10.1140/epjc/s10052-018-5947-1}{\emph{Eur. Phys.
  J. C} {\bfseries 78} (2018) 477}
  [\href{https://arxiv.org/abs/1803.07537}{{\ttfamily 1803.07537}}].

\bibitem{CMS-PAS-TOP-21-012}
{CMS Collaboration}, \emph{{Measurement of the jet mass distribution and top
  quark mass in hadronic decays of boosted top quarks in pp collisions at
  $\sqrt{s} = 13$ TeV}},  CMS Physics Analysis Summary
  \href{https://cds.cern.ch/record/2809549}{CMS-PAS-TOP-21-012}, CERN, Geneva
  (2022).

\bibitem{ATL-PHYS-PUB-2021-034}
{ATLAS Collaboration}, \emph{{A precise interpretation for the top quark mass
  parameter in ATLAS Monte Carlo simulation}},  ATLAS Public Note
  \href{https://cds.cern.ch/record/2777332}{ATL-PHYS-PUB-2021-034}, CERN,
  Geneva (2021).

\bibitem{ATLAS:2022vym}
{ATLAS Collaboration}, \emph{{Measurement of the polarisation of single top
  quarks and antiquarks produced in the $t$-channel at $\sqrt{s}=13$ TeV and
  bounds on the tWb dipole operator from the ATLAS experiment}},
  \href{https://arxiv.org/abs/2202.11382}{{\ttfamily 2202.11382}}.

\bibitem{CMS-PAS-TOP-21-014}
{CMS Collaboration}, \emph{{Measurement of the ttbar charge asymmetry in highly
  boosted events in the single-lepton channel at 13 TeV}},  CMS Physics
  Analysis Summary
  \href{https://cds.cern.ch/record/2809614}{CMS-PAS-TOP-21-014}, CERN, Geneva
  (2022).

\bibitem{ATLAS:2021dqb}
{ATLAS Collaboration}, \emph{{Measurement of the energy asymmetry in
  $t{\bar{t}}j$ production at $13\,$TeV with the ATLAS experiment and
  interpretation in the SMEFT framework}},
  \href{https://doi.org/10.1140/epjc/s10052-022-10101-w}{\emph{Eur. Phys. J. C}
  {\bfseries 82} (2022) 374}
  [\href{https://arxiv.org/abs/2110.05453}{{\ttfamily 2110.05453}}].

\end{thebibliography}\endgroup



\end{document}